\documentclass[runningheads]{llncs}

\usepackage{graphicx}
\usepackage{hyperref}
\usepackage{xcolor}
\usepackage{textcomp}
\usepackage{bbding}
\usepackage[font=scriptsize,labelfont=bf]{caption}
\usepackage{amsmath}
\usepackage{amssymb}
\usepackage{mathtools}
\usepackage{listings}
\usepackage[page]{appendix}

\hypersetup{
	colorlinks,
	linkcolor={red!50!black},
	citecolor={blue!60!black},
	urlcolor={blue!80!black},
	pdfauthor={Pegoraro Marco, Uysal Merih Seran, van der Aalst Wil M.P.},
	pdftitle={An XES Extension for Uncertain Event Data},
	pdfsubject={Process Mining over Uncertain Data},
	pdfkeywords={Event Data, Uncertainty, XES Standard, Process Mining, Business Process Management},
	pdfproducer={LaTeX},
	pdfcreator={pdfLaTeX}
}

\begin{document}

\title{An XES Extension for Uncertain Event Data\thanks{
		We thank the Alexander von Humboldt (AvH) Stiftung for supporting our research interactions. We thank and acknowledge Fabian Rempfer for his valuable input on writing style, and Majid Rafiei for his contribution to the graphics.
}
}

\author{Marco Pegoraro~\Envelope\orcidID{0000-0002-8997-7517} \and Merih Seran Uysal\orcidID{0000-0003-1115-6601} \and
	Wil M.P. van der Aalst\orcidID{0000-0002-0955-6940}}

\authorrunning{Pegoraro et al.}

\institute{Chair of Process and Data Science (PADS) \\ Department of Computer Science, RWTH Aachen University, Aachen, Germany
	\email{\{pegoraro,uysal,wvdaalst\}@pads.rwth-aachen.de}\\
	\url{http://www.pads.rwth-aachen.de/}}

\maketitle

\begin{abstract}
Event data, often stored in the form of event logs, serve as the starting point for process mining and other evidence-based process improvements. However, event data in logs are often tainted by noise, errors, and missing data. Recently, a novel body of research has emerged, with the aim to address and analyze a class of anomalies known as \emph{uncertainty}---imprecisions quantified with meta-information in the event log. This paper illustrates an extension of the XES data standard capable of representing uncertain event data. Such an extension enables input, output, and manipulation of uncertain data, as well as analysis through the process discovery and conformance checking approaches available in literature.

\end{abstract}

\section{Introduction}

Through the last decades, the increase in the availability of data generated by the execution of processes has enabled the development of the set of disciplines known as process sciences. These fields of science aim to analyze data accounting for the process perspective---the flow of events belonging to a process case.

\emph{Uncertain event data} is a newly-emerging class of anomalous event data. Uncertain data consists of events that have been logged with a quantified measure of uncertainty affecting the recorded information. Sources of uncertainty include noise, human error, or limitations of the information system supporting the process. Such imprecisions affecting the event data are either recorded in an information system with the data itself or reconstructed in a subsequent processing step, often with the aid of domain knowledge provided by process experts. Recently, the possible types of uncertain data have been classified in a taxonomy, and effective process mining algorithms for uncertain event data have been introduced~\cite{DBLP:conf/icpm/PegoraroA19,PEGORARO2021101810}. However, the data standards currently in use within the process science community do not support uncertain event logs. A very popular event data standard is XES (eXtensible Event Stream)~\cite{DBLP:conf/caise/VerbeekBDA10,van20161849}. As the name suggest, this standard has been designed to flexibly allow for extensions; in the recent past, many such extensions have been proposed, to support communications, messages and signals~\cite{leemans2017xes}, usage and performance of hardware resources~\cite{leemans2017xes2}, and privacy-preserving data transmission~\cite{DBLP:conf/bpm/RafieiA20}. This paper contributes to the field of process science by describing an XES extension which allows the representation of uncertain data, enabling XES-compatible tools to manipulate uncertain logs. Furthermore, our extension is implemented through the meta-attribute structure already supported by XES, making uncertain data retroactively readable by existing tools.

The remainder of the paper is structured as follows. Section~\ref{sec:unc} formally describes uncertain event data. Section~\ref{sec:ext} introduces an extension to the XES standard capable of representing uncertain event data. Lastly, Section~\ref{sec:conclusion} concludes the paper.

\section{Uncertain Event Data}\label{sec:unc}

In order to more clearly visualize the structure of the attributes in uncertain events, let us consider the following process instance, which is a simplified version of actually occurring anomalies, e.g., in the processes of the healthcare domain. An elderly patient enrolls in a clinical trial for an experimental treatment against myeloproliferative neoplasms, a class of blood cancers. This enrollment includes a lab exam and a visit with a specialist; then, the treatment can begin. The lab exam, performed on the 8th of July, finds a low level of platelets in the blood of the patient (event $e_2$), a condition known as thrombocytopenia (TP). During the visit on the 10th of July, the patient reports an episode of night sweats on the night of the 5th of July, prior to the lab exam (event $e_1$). The medic notes this but also hypothesizes that it might not be a symptom, since it can be caused either by the condition or by external factors (such as very warm weather). The medic also reads the medical records of the patient and sees that, shortly prior to the lab exam, the patient was undergoing a heparin treatment (a blood-thinning medication) to prevent blood clots. The thrombocytopenia, detected by the lab exam, can then be either primary (caused by the blood cancer) or secondary (caused by other factors, such as a concomitant condition). Finally, the medic finds an enlargement of the spleen (splenomegaly) in the patient (event $e_3$). It is unclear when this condition has developed: it might have appeared at any moment prior to that point. These events are collected and recorded in the trace shown in Table~\ref{table:uncertaintracestrong} within the hospital's information system.

\begin{table}[]
	\caption{The uncertain trace of an instance of healthcare process used as a running example. For the sake of clarity, we have further simplified the notation in the timestamps column by showing only the day of the month.}
	\label{table:uncertaintracestrong}
	\centering
	\begin{tabular}{ccccc}
		\textbf{Case ID}        & \textbf{Event ID} & \textbf{Timestamp}                                                                                                     & \textbf{Activity}             & \multicolumn{1}{l}{\textbf{Indeterminacy}} \\ \hline
		\multicolumn{1}{|c|}{ID192} & \multicolumn{1}{c|}{$e_1$} 
		& \multicolumn{1}{c|}{5}                                                                         & \multicolumn{1}{c|}{\emph{NightSweats}}        & \multicolumn{1}{c|}{?}                    \\ \hline
		\multicolumn{1}{|c|}{ID192}& \multicolumn{1}{c|}{$e_2$} & \multicolumn{1}{c|}{8}                                                                         & \multicolumn{1}{c|}{\emph{PrTP}, \emph{SecTP}} & \multicolumn{1}{c|}{}                    \\ \hline
		\multicolumn{1}{|c|}{ID192}& \multicolumn{1}{c|}{$e_3$} & \multicolumn{1}{c|}{4--10}                                                                         & \multicolumn{1}{c|}{\emph{Splenomeg}} & \multicolumn{1}{c|}{}                    \\ \hline
	\end{tabular}
\end{table}

In this trace, the rightmost column refers to event indeterminacy: in this case, $e_1$ has been recorded, but it might not have occurred in reality, and is marked with a ``?'' symbol. Event $e_2$ has more than one possible activity label, either \emph{PrTP} or \emph{SecTP} (primary or secondary thrombocytopenia, respectively). Lastly, event $e_3$ has an uncertain timestamp, and might have happened at any point in time between the 4th and 10th of July. These uncertain attributes do not describe the probability of the possible outcomes, and we refer to such situation as \emph{strong uncertainty}.

In some cases, uncertain events have probability values associated with them. In the example described above, suppose the medic estimates that there is a high chance (90\%) that the thrombocytopenia is primary (caused by the cancer). Furthermore, if the splenomegaly is suspected to have developed three days prior to the visit, which takes place on the 10th of July, the timestamp of event $e_3$ may be described through a Gaussian curve with $\mu = 7$. When probability is available, such attributes are affected by \emph{weak uncertainty}.

Let us now describe a data standard extension able to represent strong and weak uncertainty, enabling the analysis of uncertain data with process science techniques.

\section{An XES Standard Extension Proposal}\label{sec:ext}

The XES standard is designed to effectively represent and transfer event data, thanks to the descriptors extended from the XML language. Additionally, XES has been designed for flexibility: its descriptors, containers, and datatypes can be extended to define new types of information.

\begin{figure}[h]
	\centering
	\includegraphics[width=\textwidth, keepaspectratio]{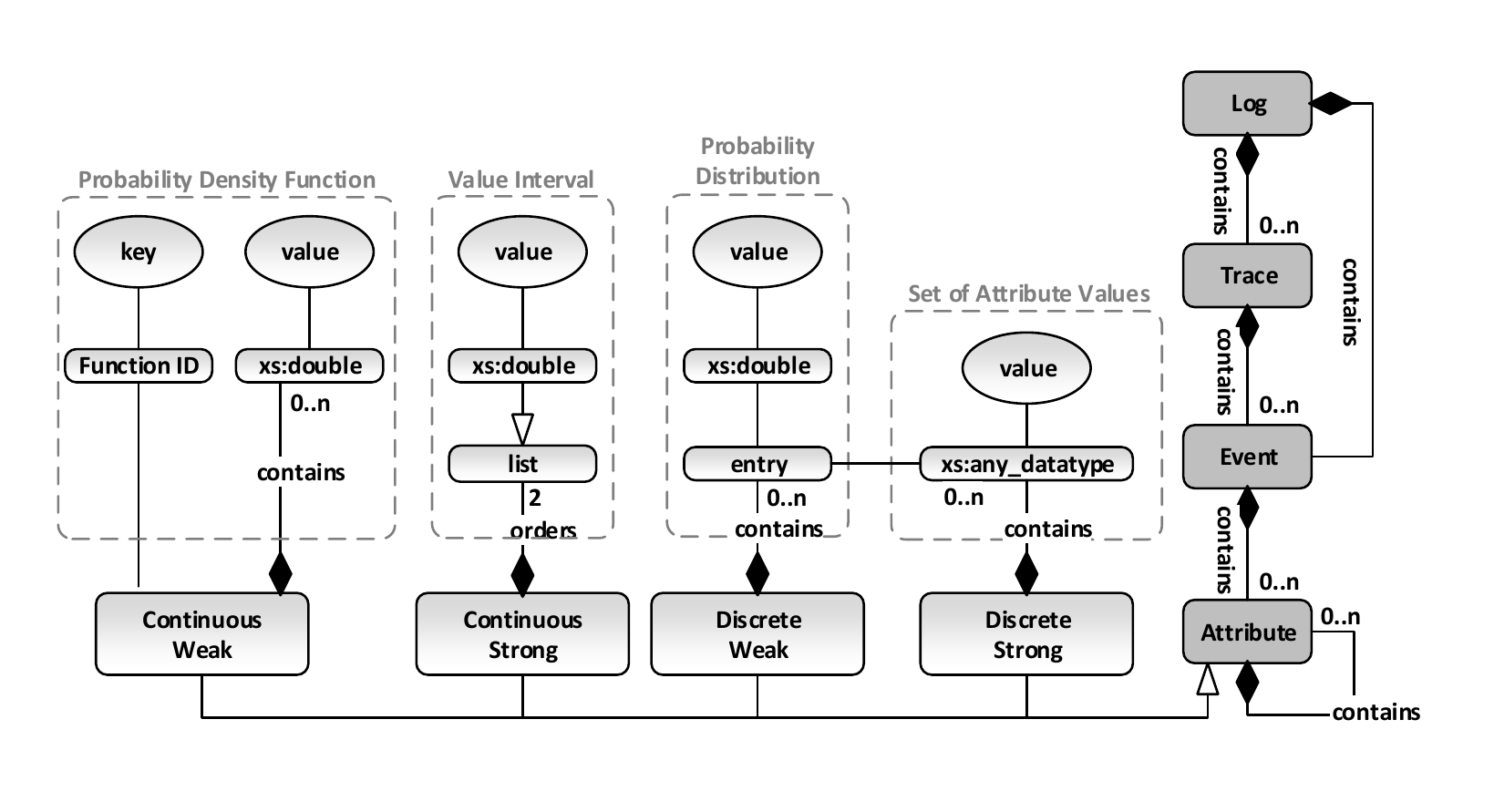}
	\caption{UML diagram illustrating an extension of the XES standard capable of representing uncertain data.}
	\label{fig:XES_extension}
\end{figure}

Figure~\ref{fig:XES_extension} describes an extension of the XES standard able to represent uncertain data as described in the previous section and illustrated in the running example of Table~\ref{table:uncertaintracestrong}.

This proposed extension can represent all scenarios of uncertain data shown in Section~\ref{sec:unc}. As a consequence, it enables XES-compliant software to import and export uncertain event data, and it allows uncertainty-aware process mining tools to implement process discovery and conformance checking approaches on uncertain data, as described in the literature.

An example of a tool able to exploit this extended XES representation to manage and analyze uncertain event data is the PROVED project\footnote{\url{https://github.com/proved-py/}}, which offers process mining and data visualization techniques capable of handling uncertain event data~\cite{10.1007/978-3-030-76983-3_24}.

It is important however to emphasize the fact that the use of the extension described here is not limited to the PROVED tool. There exist multiple tools able to support the XES standard, such as ProM~\cite{DBLP:conf/apn/DongenMVWA05}, bupaR~\cite{DBLP:conf/bpm/JanssenswillenD17}, and PM4Py~\cite{berti2019process}. Each of these tools is able to edit attributes, meta-attributes and values in a XES event log, and is then capable to record uncertain attributes on process traces. In summary, while uncertainty-aware analysis techniques are only available on a narrow selection of tools (such as PROVED), this extension benefits any tool that supports XES as one of its input/output data standards.

A set of synthetic uncertain event logs is publicly available for download\footnote{\url{https://github.com/proved-py/proved-core/tree/An_XES_Extension_for_Uncertain_Event_Data/data}}. In the same folder, it is possible to find the additional document (part of the BPM Resource track submission) explaining more in detail how our extension proposal models uncertain event data\footnote{\url{https://github.com/proved-py/proved-core/blob/An_XES_Extension_for_Uncertain_Event_Data/data/uncertainty_XES_standard.pdf}. A version of this document is reproduced in Appendix~\ref{sec:app}.}.

\section{Conclusion}\label{sec:conclusion}

Recent literature in the rapidly-growing field of process mining shows how descriptions of specific data anomalies can be extracted from information systems or obtained through domain knowledge. Anomalies labeled by such descriptions characterize uncertain event data, and there exist process mining algorithms able to exploit this meta-information to gain insights about the process with a precisely bounded reliability. A fundamental part of these data analysis approaches is however needed: formats for data representation and transmission. In this paper, we described an extension of the XES data standard which enables representation of such uncertain data, and that allows uncertain event to be read and written by existing XES-compliant software. This, in turn, empowers process mining researchers and practitioners to build analysis techniques that account for data uncertainty, and that can thus be more trustworthy and reliable.

\bibliographystyle{splncs04}
\bibliography{bibliography}

\appendix
	\section{XES Standard for Uncertain Event Data}\label{sec:app}
	In order to more clearly visualize the structure of the attributes in uncertain events, we are going to illustrate them with two examples of uncertain traces.
	
	\begin{table}[h]
		\caption{The uncertain trace of an example of healthcare process. For the sake of clarity, we have further simplified the notation in the timestamps column by showing only the day of the month.}
		\label{table:uncertaintracestrongapp}
		\centering
		\begin{tabular}{ccccc}
			\textbf{Case ID}        & \textbf{Event ID} & \textbf{Timestamp}                                                                                                     & \textbf{Activity}             & \multicolumn{1}{l}{\textbf{Indeterminacy}} \\ \hline
			\multicolumn{1}{|c|}{ID192} & \multicolumn{1}{c|}{$e_1$} 
			& \multicolumn{1}{c|}{5}                                                                         & \multicolumn{1}{c|}{\emph{NightSweats}}        & \multicolumn{1}{c|}{?}                    \\ \hline
			\multicolumn{1}{|c|}{ID192}& \multicolumn{1}{c|}{$e_2$} & \multicolumn{1}{c|}{8}                                                                         & \multicolumn{1}{c|}{\emph{PrTP}, \emph{SecTP}} & \multicolumn{1}{c|}{}                    \\ \hline
			\multicolumn{1}{|c|}{ID192}& \multicolumn{1}{c|}{$e_3$} & \multicolumn{1}{c|}{4--10}                                                                         & \multicolumn{1}{c|}{\emph{Splenomeg}} & \multicolumn{1}{c|}{}                    \\ \hline
		\end{tabular}
	\end{table}
	
	Table~\ref{table:uncertaintracestrongapp} illustrates our first example. In this trace, the rightmost column refers to event indeterminacy: in this case, $e_1$ has been recorded, but it might not have occurred in reality, and is marked with a ``?'' symbol. Event $e_2$ has more then one possible activity labels, either \emph{PrTP} or \emph{SecTP}. Lastly, event $e_3$ has an uncertain timestamp, and might have happened at any point in time between the 4th and 10th of July.
	
	In some cases, uncertain events have probability values associated with them. In the example described above, suppose the medic estimates that there is a high chance (90\%) that the thrombocytopenia is primary (caused by the cancer). Furthermore, if the splenomegaly is suspected to have developed three days prior to the visit, which takes place on the 10th of July, the timestamp of event $e_3$ may be described through a Gaussian curve with $\mu = 7$. Lastly, the probability that the event $e_1$ has been recorded but did not occur in reality may be known (for example, it may be 25\%).
	
	Assigning such probabilities to data results in the trace shown in Table~\ref{table:uncertaintraceweak}.
	
	\begin{table}[h]
		\caption{A trace where uncertain event attributes are labeled with probabilities. In this case, we also have an indeterminate event.}
		\label{table:uncertaintraceweak}
		\centering
		\begin{tabular}{ccccc}
			\textbf{Case ID}        & \textbf{Event ID} & \textbf{Timestamp}                                                                                                     & \textbf{Activity}             & \multicolumn{1}{l}{\textbf{Indeterminacy}} \\ \hline
			\multicolumn{1}{|c|}{ID348} & \multicolumn{1}{c|}{$e_4$} 
			& \multicolumn{1}{c|}{5}                                                                         & \multicolumn{1}{c|}{\emph{NightSweats}}        & \multicolumn{1}{c|}{$?: 25\%$}                    \\ \hline
			\multicolumn{1}{|c|}{ID348}& \multicolumn{1}{c|}{$e_5$} & \multicolumn{1}{c|}{8}                                                                         & \multicolumn{1}{c|}{\emph{PrTP: $90\%$}, \emph{SecTP: $10\%$}} & \multicolumn{1}{c|}{}                    \\ \hline
			\multicolumn{1}{|c|}{ID348}& \multicolumn{1}{c|}{$e_6$} & \multicolumn{1}{c|}{$\mathcal{N}(7, 1)$}                                                                         & \multicolumn{1}{c|}{\emph{Splenomeg}} & \multicolumn{1}{c|}{}                    \\ \hline
		\end{tabular}
	\end{table}
	
	Let us now formally define uncertain attributes, events, traces, and logs.
	
	\begin{definition}[Uncertain attributes]\label{def:attr}
		Let $\mathbb{U}$ be the \emph{universe of attribute domains}. Let the set $\mathcal{D} \in \mathbb{U}$ be an \emph{attribute domain}. Any $\mathcal{D} \in \mathbb{U}$ is a discrete set or a totally ordered set. A \emph{strongly uncertain attribute} of domain $\mathcal{D}$ is a subset $d \subseteq \mathcal{D}$ if $\mathcal{D}$ is a discrete set, and it is a closed interval $d = [d_{min}, d_{max}]$ with $d_{min} \in \mathcal{D}$ and $d_{max} \in \mathcal{D}$ otherwise. We denote with $S_\mathcal{D}$ the set of all such strongly uncertain attributes of domain $\mathcal{D}$. A \emph{weakly uncertain attribute} $f_\mathcal{D}$ of domain $\mathcal{D}$ is a function $f_\mathcal{D} \colon \mathcal{D} \not\to [0 ,1]$ such that $\sum_{x \in \mathcal{D}}f_\mathcal{D}(x) \leq 1$ if $\mathcal{D}$ is finite, $\int_{-\infty}^{\infty}f_\mathcal{D}(x)dx \leq 1$ otherwise. We denote with $W_\mathcal{D}$ the set of all such weakly uncertain attributes of domain $\mathcal{D}$. We collectively denote with $\mathcal{U}_\mathcal{D} = S_\mathcal{D} \cup W_\mathcal{D}$ the set of \emph{uncertain attributes} of domain $\mathcal{D}$.
	\end{definition}
	
	\begin{definition}[Uncertain events]\label{def:event}
		Let $\mathbb{U}_I$ be the \emph{universe of event identifiers}. Let $\mathbb{U}_C$ be the \emph{universe of case identifiers}.
		%Let $C \in \mathbb{U}$ be the discrete domain of all the \emph{case identifiers}.
		Let $A \in \mathbb{U}$ be the discrete domain of all the \emph{activity identifiers}. Let $T \in \mathbb{U}$ be the totally ordered domain of all the \emph{timestamp identifiers}. Let $O = \{?\} \in \mathbb{U}$, where the ``?'' symbol is a placeholder denoting \emph{event indeterminacy}. The \emph{universe of uncertain events} is denoted with $\mathcal{E} = \mathbb{U}_I \times \mathbb{U}_C \times \mathcal{U}_A \times \mathcal{U}_T \times \mathcal{U}_O$.
	\end{definition}
	
	\begin{definition}[Uncertain traces and logs]\label{def:trace_log}	
		$\sigma \subsetneq \mathcal{E}$ is an \emph{uncertain trace} if all the event identifiers in $\sigma$ are unique and all events in $\sigma$ share the same case identifier $c \in \mathbb{U}_C$. $\mathcal{T}$ denotes the universe of uncertain traces.  $L \subsetneq \mathcal{T}$ is an \emph{uncertain log} if all the event identifiers in $L$ are unique.
	\end{definition}
	
	In the notation of Definitions~\ref{def:attr},~\ref{def:event} and~\ref{def:trace_log}, the traces
	$\sigma_1$ in Table~\ref{table:uncertaintracestrongapp} and $\sigma_2$ in Table~\ref{table:uncertaintraceweak} are denoted as:
	\begin{align*}
		\sigma_1 =& \{(e_1, \text{ID192}, \{\textit{NightSweats}\}, [5, 5], \{?\}),
		\\& (e_2, \text{ID192}, \{\textit{PrTP}, \textit{SecTP}\}, [8, 8], \varnothing),
		\\& (e_3, \text{ID192}, \{\textit{Splenomeg}\}, [4, 10], \varnothing)\}
	\end{align*}
	\begin{align*}
		\sigma_2 =& \{(e_1, \text{ID348}, \{\textit{NightSweats}\}, [5, 5], \{(?, 0.25)\}),
		\\& (e_2, \text{ID348}, \{(\textit{PrTP}, 0.85), (\textit{SecTP}, 0.15)\}, [8, 8], \varnothing),
		\\& (e_3, \text{ID348}, \{\textit{Splenomeg}\}, \mathcal{N}(7, 1), \varnothing)\}
	\end{align*}
	The attribute domains are\footnote{Here, we defined the timestamp domain as the set $\mathbb{N}$ of natural numbers. The usual mathematical notation is unwieldy and unsuitable to represent complete timestamps as normally read and represented by humans; however, it is easy to see how a precise date and time can be represented by an integer without loss of information through conventions such as the Unix time (seconds since the Epoch, or fractions thereof).}:
	\begin{align*}
		A =& \{NightSweats, PrTP, SecTP, Splenomeg\}
		\\T =& \mathbb{N}
		\\O =& \{?\}
	\end{align*}
	Examples of uncertain attributes are:
	\begin{align*}
		S_A =& \{\{\textit{PrTP}, \textit{SecTP}\}, \{\textit{NightSweats}, \textit{PrTP}\}, \{\textit{Splenomeg}, \textit{PrTP}, \textit{SecTP}\}, \dots\}
		\\S_T =& \{[5, 5], [8, 8], [4, 10], [1, 1], [10, 12], [10, 16], \dots\}
		\\S_O =& \{\varnothing, \{?\}\}
		\\W_A =& \{\{(\textit{PrTP}, 0.85), (\textit{SecTP}, 0.15)\}, \{(\textit{NightSweats}, 0.90)\},\\& \{(\textit{Splenomeg}, 0.70), (\textit{PrTP}, 0.20), (\textit{SecTP}, 0.10)\}, \dots\}
		\\W_T =& \{\mathcal{N}(7, 1), U(4, 10), \Gamma(3, 2), \dots\}
		\\W_O =& \{\{(?, 0.25)\}, \{(?, 0.05)\}, \{(?, 0.90)\}, \dots\}
	\end{align*}
	
	Note that, while the most usual case would involve label attribute values with a complete probability distribution (probabilities summing to 1), here we allow for a sum $\leq 1$, to enable maximum flexibility in uncertain data representation.
	
	This mathematical framework allows to represent events with uncertain attributes, both strongly and weakly uncertain, and both in the discrete and continuous domains. We will now see how to represent such events in the XES standard.
	
	In this extension, discrete strongly uncertain attributes are represented by a container of data with any type: this represents a set of arbitrary objects, which are the possible values of the uncertain attribute. In the totally ordered case, the uncertain attribute is modeled by a list of two sorted values. Such values represent the extremes of an interval in which the values of the uncertain attribute can range. The following code snippet contains the full representation of the trace in Table~\ref{table:uncertaintracestrong}.
	
	\definecolor{light-gray}{gray}{0.85}
	\lstset{
		numbers=left,
		breaklines=true,
		frame=single,
		tabsize=1,
		basicstyle=\ttfamily,
		literate={\ \ }{{\ }}1
	}
	\begin{lstlisting}[basicstyle=\scriptsize,language=XML]
	<trace>
	<string key="concept:name" value="ID192"/>
	<event>
	<string key="concept:name" value="NightSweats"/>
	<date key="time:timestamp" value="2011-07-05T12:00:00+00:00"/>
	<container key="uncertainty:discrete_strong">
	<bool key="uncertainty:indeterminacy" value="true"/>
	</container>
	</event>
	<event>
	<string key="concept:name" value="PrTP"/>
	<date key="time:timestamp" value="2011-07-08T12:00:00+00:00"/>
	<container key="uncertainty:discrete_strong">
	<string key="concept:name" value="PrTP"/>
	<string key="concept:name" value="SecTP"/>
	</container>
	</event>
	<event>
	<string key="concept:name" value="Splenomeg"/>
	<date key="time:timestamp" value="2011-07-07T12:00:00+00:00"/>
	<list key="uncertainty:continuous_strong">
	<date key="time:timestamp" value="2011-07-04T12:00:00+00:00"/>
	<date key="time:timestamp" value="2011-07-10T12:00:00+00:00"/>
	</list>
	</event>
	</trace>
	\end{lstlisting}
	
	Weak uncertainty is also modeled by our extension. In this scenario, the discrete attributes are represented by a container of \texttt{uncertainty:entry} objects, which are pairs constituted by an attribute value and its probability. Lastly, the totally ordered case is described by a probability function, which is identified by a key and a set of parameters. We can see an example of these in the representation of the trace in Table~\ref{table:uncertaintraceweak}, contained in the following code snippet.
	
	\begin{lstlisting}[basicstyle=\scriptsize,language=XML]
	<trace>
	<string key="concept:name" value="ID192"/>
	<event>
	<string key="concept:name" value="NightSweats"/>
	<date key="time:timestamp" value="2011-07-05T12:00:00+00:00"/>
	<container key="uncertainty:discrete_weak">
	<container key="uncertainty:entry">
	<bool key="uncertainty:indeterminacy" value="true"/>
	<double key="uncertainty:probability" value="0.25"/>
	</container>
	</container>
	</event>
	<event>
	<string key="concept:name" value="PrTP"/>
	<date key="time:timestamp" value="2011-07-08T12:00:00+00:00"/>
	<container key="uncertainty:discrete_weak">
	<container key="uncertainty:entry">
	<string key="concept:name" value="PrTP"/>
	<double key="uncertainty:probability" value="0.90"/>
	</container>
	<container key="uncertainty:entry">
	<string key="concept:name" value="SecTP"/>
	<double key="uncertainty:probability" value="0.10"/>
	</container>
	</container>
	</event>
	<event>
	<string key="concept:name" value="Splenomeg"/>
	<date key="time:timestamp" value="2011-07-07T12:00:00+00:00"/>
	<container key="uncertainty:continuous_weak">
	<string key="uncertainty:density_function" value="GAUSSIAN"/>
	<list key="uncertainty:function_parameters">
	<double key="parameter_mean" value="7"/>
	<double key="parameter_stddev" value="1"/>
	</list>
	</container>
	</event>
	</trace>
	\end{lstlisting}
	
	A set of synthetic uncertain event logs is publicly available for download\footnote{\url{https://github.com/proved-py/proved-core/tree/An_XES_Extension_for_Uncertain_Event_Data/data}}.

\end{document}